\documentclass[
    ,final            
  ]
  {aipproc}

\layoutstyle{6x9}

\def\Swift{\emph{Swift}}

\def\til{\ensuremath{\sim\,}}

\def\s{\ensuremath{\sigma}}

\begin{document}

\title{Improving Swift-XRT positions of GRBs}


\author{P.A. Evans}{address={Department of Physics and Astronomy, University of Leicester, University Road, Leicester, LE1 7RH, UK.}}
\author{A.P. Beardmore}{address={Department of Physics and Astronomy, University of Leicester, University Road, Leicester, LE1 7RH, UK.}}
\author{M.R. Goad}{address={Department of Physics and Astronomy, University of Leicester, University Road, Leicester, LE1 7RH, UK.}}
\author{J.P. Osborne}{address={Department of Physics and Astronomy, University of Leicester, University Road, Leicester, LE1 7RH, UK.}}
\author{D.N. Burrows}{address={Department of Astronomy and Astrophysics, 525 Davey Lab, Pennysylvania State University, University Park, PA 16802, USA}}
\author{N. Gehrels}{address={NASA/Goddard Space Flight Center, Greenbelt, MD 20771, USA}}

\begin{abstract}
Since GRBs fade rapidly, it is important to publish accurate, precise positions
at early times. For \Swift-detected bursts, the best promptly available position
is most commonly the X-ray Telescope (XRT) position. We present two processes, 
developed by the \Swift\ team at Leicester, which are now routinely used to
improve the precision and accuracy of the XRT positions reported by the \Swift\ 
team. Both methods, which are fully automated, make use of a PSF-fitting
approach which accounts for the bad columns on the CCD. The first method yields
positions with 90\% error radii $<$4.4" 90\% of the time, within 10--20 minutes
of the trigger. The second method astrometrically corrects the position using
UVOT field stars and the known mapping between the XRT and UVOT detectors,
yielding enhanced positions with 90\% error radii of $<$2.8" 90\% of the time,
usually ~2 hours after the trigger.
\end{abstract}

\maketitle


\section{Introduction}
For the majority of \Swift-detected GRBs, the best promptly available position
is that of the X-ray telescope (XRT, \cite{Burrows05}). It is thus desirable to
reduce the 3.5" boresight uncertainty associated with this instrument.

\begin{figure}
\includegraphics[height=8cm,angle=-90]{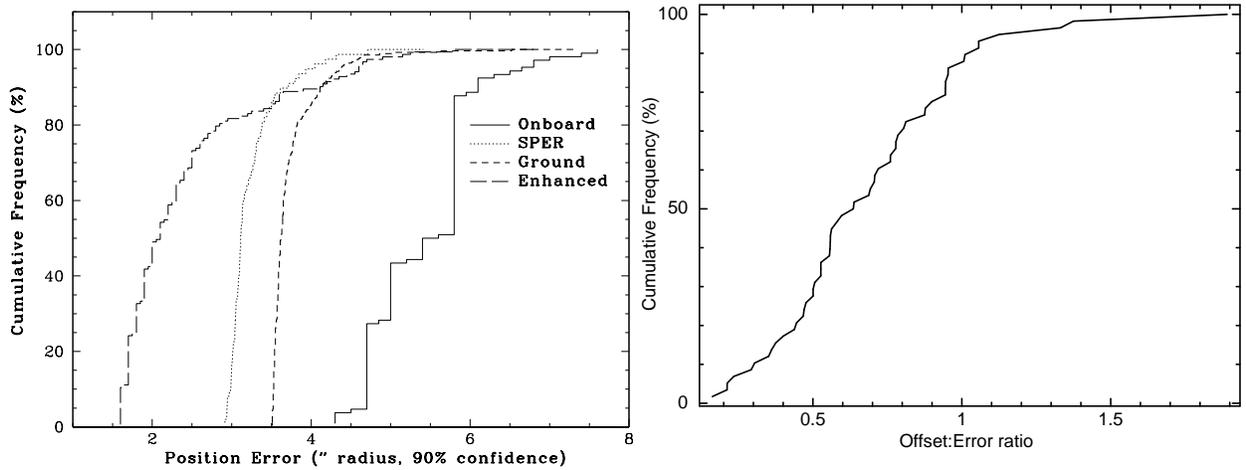}
\includegraphics[height=8cm,angle=-90]{errcheck}
\caption{\emph{Left:} Distribution of the 90\% confidence error radii produced
for XRT positions of GRBs. Distributions shows are those obtained onboard
automatically (solid), from SPER data (Section 2, dotted), on the ground from
the full dataset (short dashes) and the enhanced positions (Section 3, long
dashes).\emph{Right:} Distribution of the offsets of the UVOT-enhanced positions
from the UVOT position for GRBs with both, divided by the position error. As can
be seen, 90\% of the enhanced positions agree with the UVOT positions,
confirming the error circle is correctly calibrated.}
\label{fig:errdist}
\end{figure}

We have developed two techniques to achieve this goal. The first is a fitting
technique which accounts for hot columns on the X-ray CCD. This is described in
Section~1 and the application of this to promptly available data is detailed in
Section~2. The second technique is applied to the full ground dataset, and uses
the field stars in the UV/Optical telescope (UVOT, \cite{Roming05}) to
astrometrically correct the XRT position, eliminating the XRT's boresight
uncertainty. This is described in Section~3. In Fig.~\ref{fig:errdist} we show
the distribution of position uncertainties produced by these techniques,
comparing them with positions determined onboard the XRT, and the `refined'
positions produced from the full dataset without astrometric correction.
Finally, in Section~4 we discuss forthcoming improvements to the second
technique, and the potential for applying it to the prompt data. For an overview
of the different positions available from the \Swift\ XRT, see
\url{http://www.swift.ac.uk/xrt_pos.php}

\section{1. PSF fitting}
\label{sec:psf}
Given an XRT image, we first apply a cell-detect routine to locate sources and
provide approximate positions. Thereafter, following \cite{Cash78} we fit the
Point Spread Function (PSF) of each source with the theoretical XRT PSF
(\cite{Moretti05}), using two free parameters -- the $x$ and $y$ position of the
object. This fit is performed in CCD detector coordinates so that the positions
of the hot columns are known and the model PSF normalisation can be adjusted
accordingly. Note that this fit is not used to calculate the onboard or `refined' positions.

\begin{figure}
\includegraphics[width=10cm]{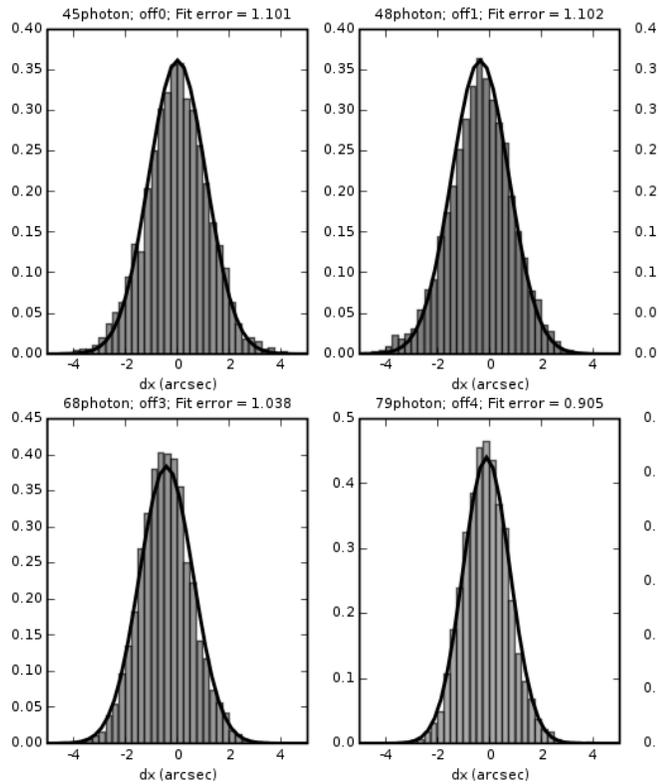}
\caption{Results of fitting 5000 simulated images(per panel) with 100 photons
per image. In the upper left panel the source is centred on the bad columns, and
then it moves away from them one pixel at a time from the left to right panels, then top
to bottom.}
\label{fig:psftest}
\end{figure}

We tested this by simulating images where the real object position is known. We
performed this simulation 5000 times for a given object position, and applied
the PSF fit to each image. Fig.~\ref{fig:psftest} shows the results for a range
of positions starting over the bad columns and moving away from them. The
histograms show the distance from the fitted position to the real position, and
the solid line shows a Gaussian with a \s\ corresponding to the typical fit
error. This shows both that the reported error is accurate, and that the fit
performs well even when the object lies on the bad columns.

\section{2. Prompt Positions}
\label{sec:sper}
During the first snapshot after a GRB trigger, single pixel Photon Counting (PC)
mode X-ray events with $E>0.5$ keV are telemetered to the ground via TDRSS.
These Single Pixel Event Report (SPER) packages are distributed to the \Swift\ XRT
team every 390 seconds, an image is automatically extracted and the PSF fit
(Section~1) is applied. Comparing positions thus produced with UVOT positions
of optical counterparts, we find that the systematic uncertainty associated with
the XRT boresight can be reduced to 2.9", from the 3.5" deduced using a
barycentric fit, thus these positions offer an improvement over the previous
standard.

As SPER data are telemetered every 390 s, there are usually multiple deliveries
per GRB. Experience shows that the position can be substantially improved
between the first and second deliveries, but only minimally thereafter. Thus,
either when the second SPER has been processed, or 9 minutes after the first
SPER was received (if no second one has arrived), the position is distributed as
a GCN Position (Update) Notice. This position is used in the initial GCN
Circular prepared by the \Swift\ team, which details the detection of the burst.
All SPER positions are published online as soon as they are produced, at
\url{http://www.swift.ac.uk/spertable.php}.

\section{3. Enhanced Positions}
\label{sec:enh}
As noted above, standard XRT positions have a systematic error of 3.5",
theoretically limiting the XRT's position accuracy to this value. However, by
determining the mapping between the XRT and UVOT detectors, we are able to use
UVOT field astrometry to correct the XRT boresight, reducing the systematic to
1.5". Note that this process does not require UVOT to detect the GRB, and works
for \til70\% of \Swift-detected bursts.

Full details of this process are given in \cite{Goad07}. A summary is: Use the
PSF fit described above to determine the XRT detector position of a GRB, convert
this into an equivalent UVOT detector position, and hence UVOT sky position.
Align the UVOT field of view with the USNO-B1 catalogue to correct this
position.

Because \Swift\ does not remain perfectly steady, the XRT detector position of
an object can drift during a snapshot, meaning we can only use times of
simultaneous X-ray and UVOT data. Further, we limited the current version of our
software to the UVOT $V$ filter, as it was for this that the map was determined.
Data are thus split into \emph{overlaps}, of simultaneous XRT PC mode data and UVOT $V$-band
data. The above process is applied to each overlap in turn, yielding one
position per overlap. The weighted mean of these is then calculated, any individual
positions more than $3\s$ from this are discarded and the mean is recalculated.
Finally, the 1.5" systematic error arising from uncertainty in the UVOT-XRT mapping is
added in quadrature to give the UVOT-enhanced XRT position. The position is
immediately circulated to the XRT team, and posted online at
\url{http://www.swift.ac.uk/xrt_positions}. When a position is first determined for a
GRB, it is also distributed to the community in an automatically generated GCN
circular (see \cite{Osborne07}). Note that, if the UVOT-enhanced position is the
first X-ray position found for a GRB, the automatic circular will not be sent.
This is to give the XRT team a chance to check that the (probably faint) source
is the afterglow. The position will still be posted online, and distributed via
a circular when verified.

\section{4. Future improvements}
\label{sec:future}
We are currently developing a second version of the UVOT-enhancement software.
This code makes use of multiple UVOT filters ($V$, $B$ and white), multiple
\Swift\ obsIDs, and an improved PSF fit algorithm which works in sky
co-ordinates, using exposure maps to correct for bad columns, bad pixels and
vignetting. This also allows for non-simultaneous X-ray and UVOT data to be
used. Tests suggest that the new version reduces typical total error radii by
25--50\%. Once testing is complete, the new version will take over live
processing of GRBs -- this will be announced via a GCN circular, hopefully in
early 2008.

A parallel development is that, using the new version of the code, we are able
to use the limited data products available immediately after a GRB trigger to
UVOT-enhance the SPER data, combining the methods of Sections~2 and~3 in this paper.
The improvement in SPER positions is less pronounced than with the full dataset,
however we anticipate a reduction in error radius of \til25\% for a typical GRB.




\bibliographystyle{aipproc}   

\bibliography{evans}

\IfFileExists{\jobname.bbl}{}
 {\typeout{}
  \typeout{******************************************}
  \typeout{** Please run "bibtex \jobname" to optain}
  \typeout{** the bibliography and then re-run LaTeX}
  \typeout{** twice to fix the references!}
  \typeout{******************************************}
  \typeout{}
 }

\end{document}